\documentclass{article}
\usepackage[utf8]{inputenc}
\usepackage{graphicx}
\usepackage{times}
\usepackage{caption}
\usepackage{subcaption}
\usepackage{epstopdf} 

\date{}

\author{Alfredo J. Morales*, Shahar Somin, Yaniv Altshuler \\ and Alex 'Sandy' Pentland\\
MIT Media Lab\\ alfredom@mit.edu}
\title{User behavior and token adoption on ERC20}

\begin{document}
\maketitle

\begin{abstract}
Cryptocurrencies and Blockchain-based technologies are disrupting all markets. While the potential of such technologies remains to be seen, there is a current need to understand emergent patterns of user behavior and token adoption in order to design future products. In this paper we analyze the social dynamics taking place during one arbitrary day on the ERC20 platform. We characterize the network of token transactions among agents. We show heterogeneous profiles of user behavior, portfolio diversity, and token adoption. While most users are specialized in transacting with a few tokens, those that have diverse portfolios are bridging across large parts of the network and may jeopardize the system stability. We believe this work to be a foundation for unveiling the usage dynamics of crypto-currencies networks.
\end{abstract}

\section{Introduction}

During the last years we have experienced an explosion of applications and solutions based on Blockchain technologies  \cite{nakamoto2008bitcoin,Swan:2015:BBN:3006358}. Bitcoin and other platforms such as Ethereum have implemented and established decentralized ledgers to perform transactions using digital currencies and tokens worldwide. On these platforms, transactions occur through a distributed system that validates operations by making collective decisions rather than querying centralized authorities. While many researchers and innovators are currently working on creating new applications for this type of technologies \cite{WEF2018}, it is a current challenge for product developers to understand how people behave, interact and self-organize in such systems. Human behavioral patterns reveal choices and preferences \cite{Pentland:2008:HST:1450928} which can inform product development models \cite{clack2016smart} for solving critical issues such as token adoption and valuation \cite{tokcatalini2018initial,tokyoder1983price,tokcong2018tokenomics}.

Blockchain based solutions can be understood in the framework of multi-layered systems \cite{DBLP:journals/corr/abs-1807-00955,PhysRevE.89.032804}. A web of computers at the bottom layer compete with each other for validating transactions and storing them on the chain. This process is called {\it mining} and comprises the technical side of the platform. Mining is the mechanism that ensures secure decentralized operations \cite{8029379}. Miners can be rewarded for validating transactions. Transactions are first validated locally against a copy of the ledger and later contrasted with results from other miners before final inclusion on the chain. The security \cite{Eyal:2018:MEB:3234519.3212998}, privacy \cite{rahulamathavan2017privacy} and scalability \cite{eyal2016bitcoin} of these tasks are major technological challenges for the future development of tokenized systems.

Another network runs on top of miners which is a social network \cite{yaniv2018,7796940,fi8010007}. In this network, users (and robots) perform transactions with each other. These transactions can consist of money transfer or other applications depending on the token and associated service. Understanding the structure and dynamics of social networks \cite{Krafft2016,Yaniv2014,Pentland2012}, as well as user behaviors more generally, is crucial for designing new products and evaluating the performance of current ones  \cite{UserBeh2006}. The complexity of possible behaviors is measured in terms of information \cite{BarYam2005ComplexityR} and capturing it is fundamental for building strategies that deal with changing environments \cite{Ashby1991}. 

ERC20 is a protocol to implement tokens on Ethereum--the platform of the ETH crypto-currency. Tokens are created with smart contracts \cite{Pompianu2017,Anderson2016}, which consist in software that runs when transactions are performed. Transactions are based on tokens, which are assigned to specific contracts. Users can either buy ETH or tokens of other contracts. Not all contracts are open to the public. Some of them are closed and private. Analogously, not all tokens have an economic purpose \cite{Christidis2016}. Tokens can symbolize company shares, property documents, or supply chain steps, among others.

Tokens may revolutionize economic activities. Current developments include new ways to raise funds for projects and entrepreneurs, as well as new ways to deploy and monetize digital applications for innovators \cite{CHEN2018567}. Different economic activities may tune the currency they operate with in order to satisfy their particular needs \cite{DBLP:journals/corr/abs-1807-00955}. 
Some activities require tokens to be adopted in order to interact with the service, while other tokens can be simply acquired and held in the wallet. Understanding the reasons for token adoption and use is a major challenge for Blockchain product developers. The large variety of use purposes and services makes it non-trivial to guess the value and potential adoption of tokens. Access to data provides the opportunity to measure and observe the current use of Blockchain services and learn from their statistical properties.

In this work we study the social network that arises during transactions on the ERC20 platform during an arbitrary day. In a previous work \cite{yaniv20182}, we showed that the transaction network reached a stable, complex topology over time \cite{Pentland2014}. That study validated the application of network analysis for understanding the structure of tokenized networks. Here we further investigate the complex structure of the transaction network and characterize token adoption and transaction activity. We show heterogeneous user roles and behaviors in terms of transaction diversity and activity. We also characterize tokens in terms of emergent adoption and properties in the transaction network.

In section \ref{methods} we present our data set and describe how we construct the networks. In section \ref{results} we show visualizations and properties of the transaction network, including those of users and tokens. In section \ref{discussion} we discuss our results and provide conclusions.

\section{Methods and Data}
\label{methods}

We use transaction data from the ERC20 platform on Nov 4th, 2018. Each record represents a transaction. It includes a time stamp and identifiers for the seller, buyer and token. Identifiers are formatted by hashcodes. The dataset has a total of 150,506 records, regarding 2,718 tokens and 86,810 users. Out of the total, 52,446 users are exclusive buyers, 15,398 users are exclusive sellers and 18,966 users do both activities.

In order to analyze transaction activity on ERC20, we create a social network. Users are represented as nodes. Edges represent token transactions. Edges are directed from user $i$ (seller) to $j$ (buyer) and weighted by the total number of transactions between $i$ and $j$. Each transaction is associated with a token. Therefore, edges also represent the set of tokens exchanged between $i$ and $j$. 


\section{Results}
\label{results}

\begin{figure}
    \centering
  \includegraphics[width=\linewidth]{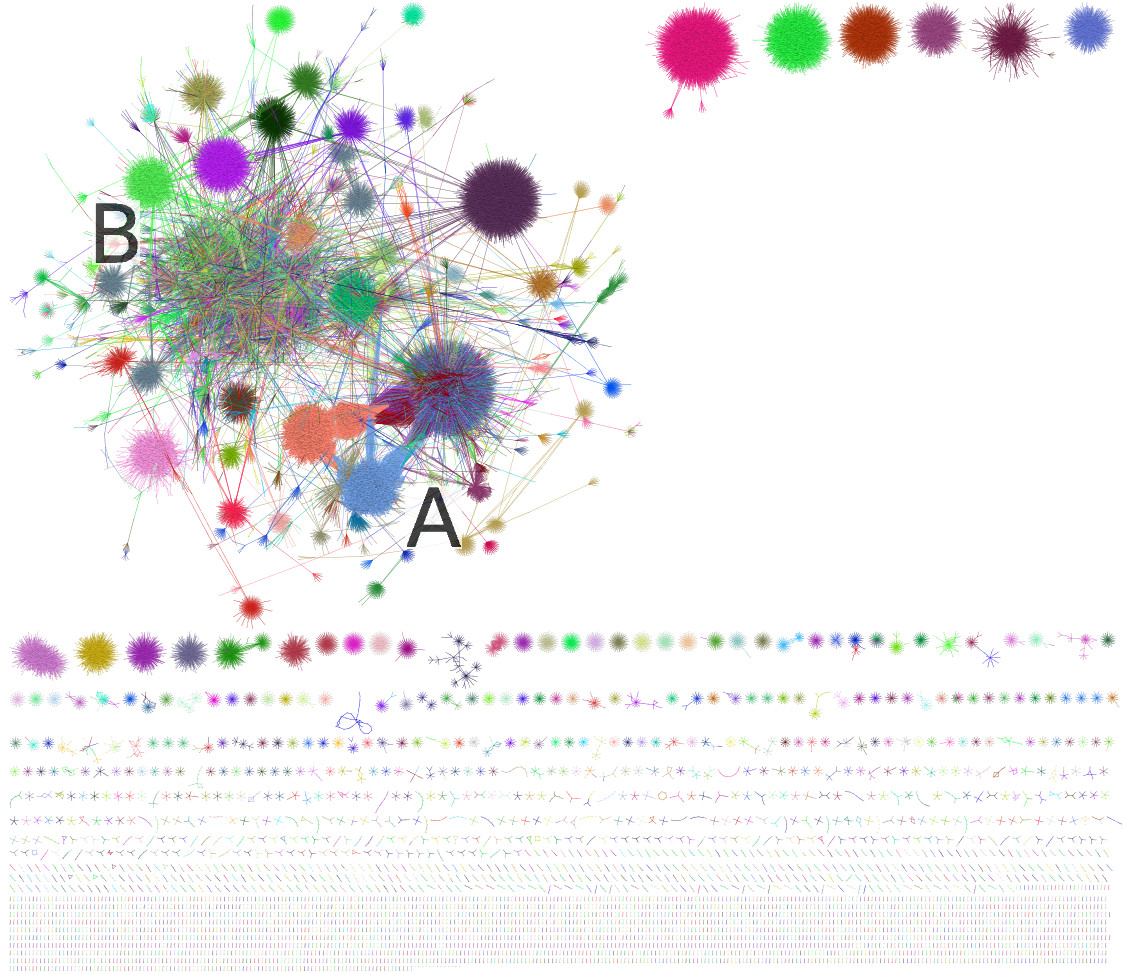}
  \caption{ERC20 transaction network during an arbitrary day. Nodes represent users, edges represent transactions and colors indicate the token being transacted. There are multiple disconnected components associated to different tokens and a giant component shown in the upper left. (A) shows a region of three dominant clusters and (B) shows a region of high token mixing.}
  \label{fig:visualization}
\end{figure}

\begin{figure}
\centering
  \includegraphics[width=0.8\linewidth]{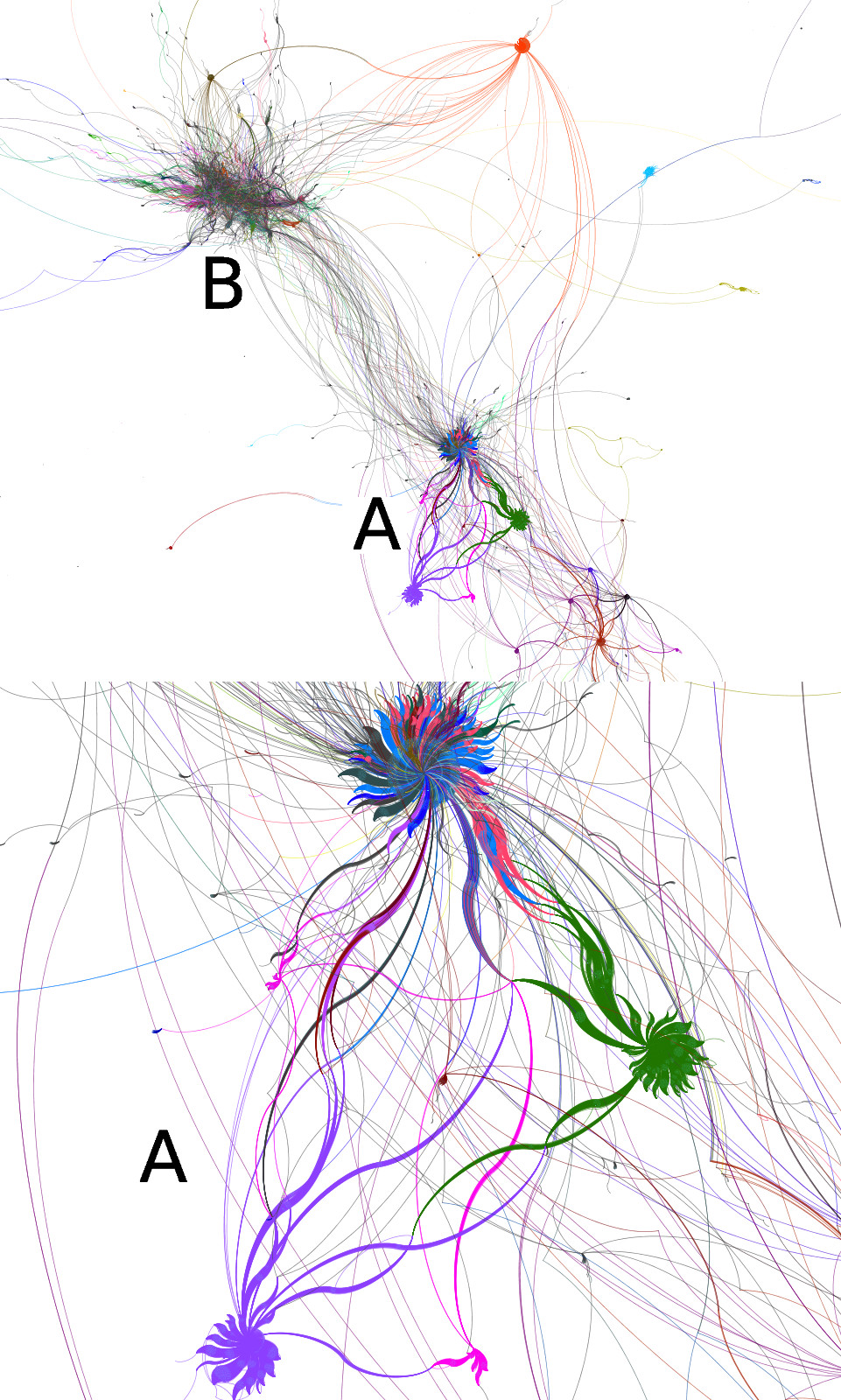}
  \caption{Giant component of ERC20 transaction network during an arbitrary day. Nodes represent users, edges represent transactions and colors indicate the token being transacted.  (A) shows a region of three dominant clusters and (B) shows a region of high token mixing.  In the bottom panel, two of the influential clusters trade with a specific toke, while the other one trades with multiple ones. Hubs and tokens are usually connected by several hops. The edge extension is shown by the direction of the arc.}
  \label{fig:visualization_sub}
\end{figure}

Visualizations of the transaction network are presented in Figures \ref{fig:visualization} (full network) and \ref{fig:visualization_sub} (giant component). Users are represented by dots and connections indicate transactions. Distinct edge colors represent transacted tokens. If a couple users exchange more than one token, the edge is colored according to the mostly transacted one. The structure of this network is comprised by a set of disconnected components (Figure \ref{fig:visualization} bottom) and one giant component (Figure \ref{fig:visualization} top left and Figure  \ref{fig:visualization_sub}). The disconnected components are usually associated to a single token. Their adoption size ranges widely, from a couple or few users (located at the bottom of Figure \ref{fig:visualization})  up to several hundreds (located at the top of Figure \ref{fig:visualization}). These colored modules are respectively paired to smart contracts and tokenized services. Network representations inform about structural patterns of token adoption beyond the total number of adopters.

The larger disconnected components in Figure \ref{fig:visualization} present a hub-spoke structure. They respectively consist of a few sellers per tokens (two in a few cases) transacting with multiple buyers exclusively. By sorting components by number of users, the figure shows the progression of token adoption from a single transaction up to the hub-spoke structures. Some tokens are aggregated with each other in the giant component as intermediary users trade and bridge across communities. The properties of the giant component network are presented in Table \ref {tab:gc_properties}. The small world effect is present, and clustering is low (the network is very sparse). The degree assortativity is negative, indicating that hubs are usually connected to lowly connected users rather than to other hubs. 

\begin{table}[t]
    \centering
\begin{tabular}{ |c|c| }
 \hline
 Property & Value \\
 \hline
 Average shortest path length & 6.23 \\
 Average clustering  & 0.03 \\ 
 Degree assortativity & -0.2 \\ 
 \hline
\end{tabular}
 \caption{Network properties of the ERC20 transaction network giant component.}
\label{tab:gc_properties}
\end{table}

The giant component presents two dominant behaviors labeled A and B in Figures \ref{fig:visualization} and \ref{fig:visualization_sub}. The region A of the giant component is dominated by a few hubs whose tokens represent an important share of the transactions (Figure \ref{fig:visualization}A and \ref{fig:visualization_sub}A). Hubs are not directly connected to each other. Instead, a chain of intermediate users connect hubs. For example, 3 to 4 users connect the purple and green hubs shown in Figure \ref{fig:visualization_sub} (bottom panel). On both sides, the first hops transact the same token as the hubs, and only an intermediate couple users exchange the two different tokens. The extension of tokens across the network follows an exponential distribution with an average of 2.48 hops and a maximum of 19 hops.

The region B of the giant component in Figures \ref{fig:visualization} and \ref{fig:visualization_sub} presents a different behavior than the hub-spoke structures observed in region A. In region B, users are densely connected to each other through a diverse variety of small tokens. This diverse region is also shown in the giant component visualization in Figure \ref{fig:visualization_sub}B. In this visualization, tokens representing less than 0.1\% are colored in black and the diverse region is shown as an entangled black sub-network in the upper left corner.

\begin{figure}
   \includegraphics[width=\linewidth]{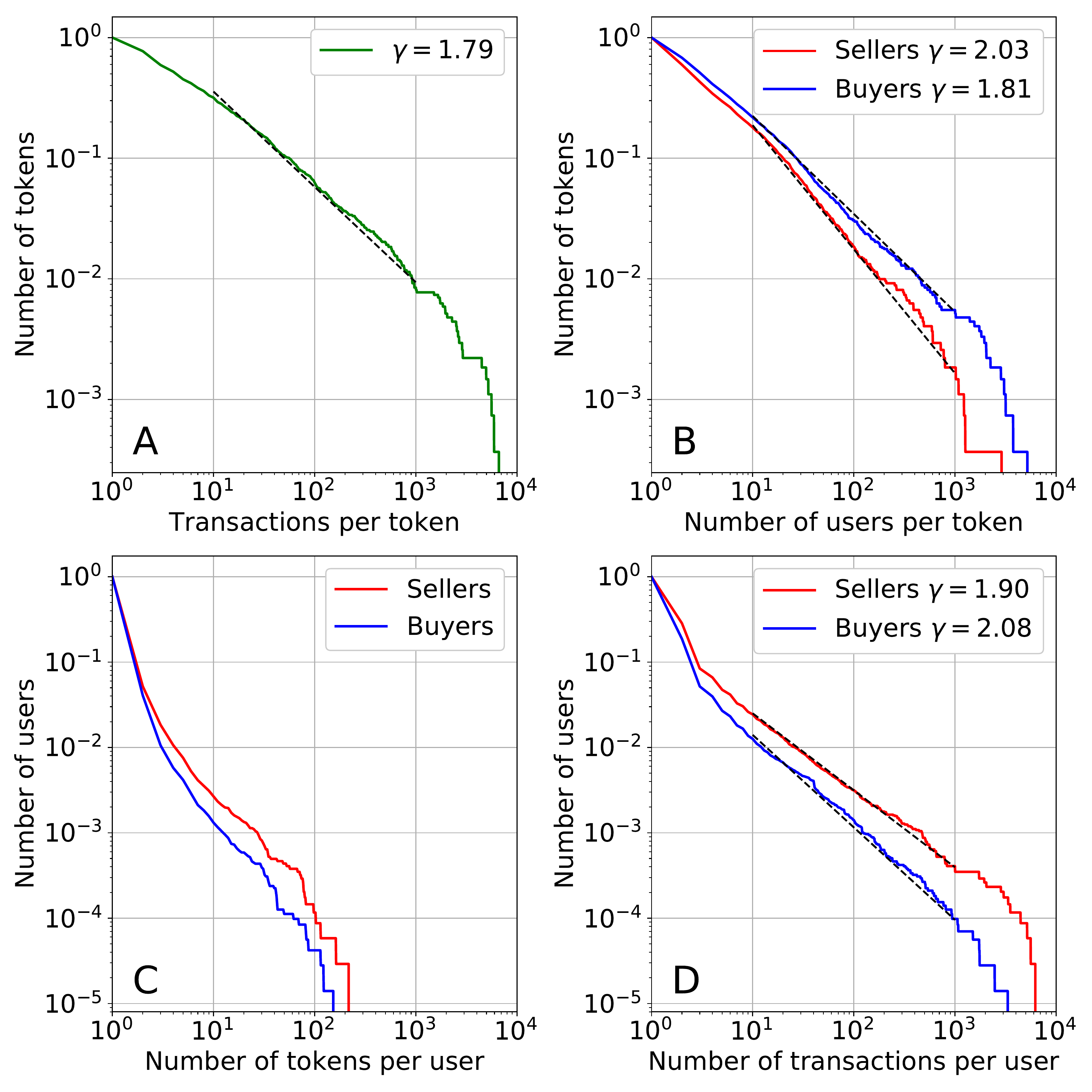}
  \caption{Token adoption statistical behavior. Complementary cumulative distribution (CCDF) of (A) the number of transactions per token, (B) the number of tokens transacted per user, (C) number of users per tokes, (D) the number of transactions per user. Token sellers are represented in red and buyers in blue. The black dashed lines show the fitted curve: $f(x) = a x^{-(\gamma-1)}$. $x \in (10,1000)$ in panels A.
  Respective $\gamma$ values are shown in the legend.  }
  \label{fig:distributions}
\end{figure}

The behavior of token adoption is highly heterogeneous. While some tokens are widely exchanged and belong to the giant component shown in Figure \ref{fig:visualization}, others are isolated and hardly used. The complementary cumulative distribution of transactions per token is presented in Figure \ref{fig:distributions}A. The distribution follows a power law with exponent $\gamma=1.79$. The power law distribution indicates that while most of tokens are transacted only a couple times, there is a minority of highly popular tokens that are widely adopted. Those tokens dominate the network and concentrate the large majority of transactions.

Most users buy or sell just a couple of tokens while a very few of them transact with multiple ones-- up to hundreds. The complementary cumulative distribution (CCDF) of tokens by user is presented in Figure \ref{fig:distributions}C, for both buyers (blue) and sellers (red) respectively. Most users are specialized in trading with a given token. A minority of users buy and sell with diverse portfolios. A regression analysis of the diversity of portfolios as a function of network properties is presented in Table \ref{tab:regression} and discussed below.

There are more buyers than sellers in the dataset. While most tokens are either sold or bought by a very few users, there is a set of tokens that are adopted by the large majority of users. The complementary cumulative distribution of users per token is presented in Figure \ref{fig:distributions}B. The blue curve shows the number of buyers per token and the red curve shows the number of sellers. These distributions also follow a power law behavior across multiple orders of magnitude. In this case the tail of buyers ($\gamma=1.81$) decays slower than the tail of sellers ($\gamma=2.03$). That means that the number of tokens with a large variety of buyers is higher than the number of tokens with a large variety of sellers.

Sellers are fewer yet concentrate most transactions. The complementary cumulative distribution of transactions per user is presented in Figure \ref{fig:distributions}D for both buyers (blue) and sellers (red). Both distributions follow a power law behavior across multiple orders of magnitude ($\gamma=2.08$ for buyers and $\gamma=1.90$ for sellers). This indicates that while most users sell or buy just a couple of tokens at most, there is a minority of highly active users who account for most of the transactions. These users are on the tail of the distributions and more frequent in the case of sellers than buyers. While tokens have more buyers than sellers, the latter concentrate more transactions.

\begin{figure}
  \includegraphics[width=\linewidth]{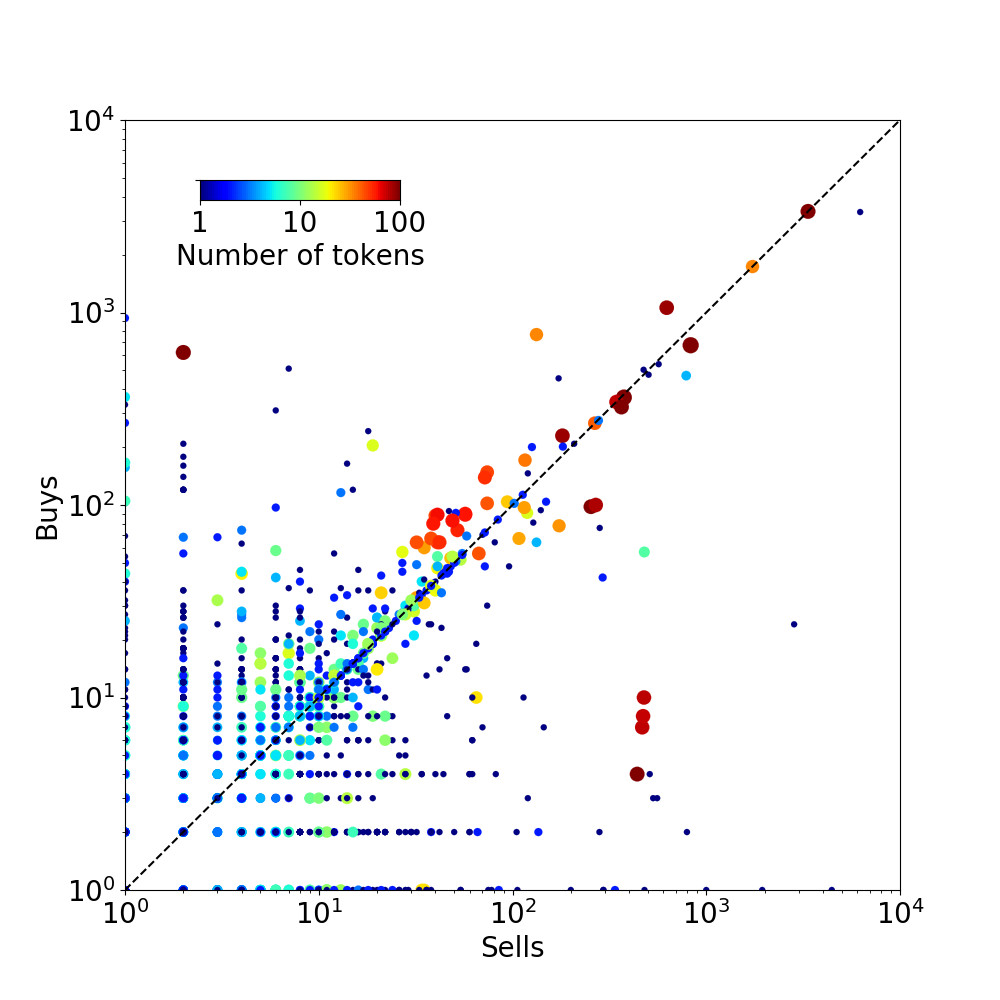}
  \caption{Scatter plot of the number of sells and buys per user. Dots indicate users. Color and size are proportional to the number of tokens transacted by user. Scale on figure.}
  \label{fig:transactions_per_user_scatter}
\end{figure}

In order to understand the behavior of buyers and sellers and their token adoption, we present a scatter plot with the number of buys and sells per user in Figure \ref{fig:transactions_per_user_scatter}. Dots indicate users and color indicates the number of transacted tokens. There are three dominant user behaviors: buyers, sellers and both (shown in the dot location); as well as two types of token portfolios: specialized and diversified (shown in the dot color). Users located above the diagonal mostly buy tokens, while users located below the diagonal mostly sell them. These users are predominantly specialized in a given token (blue dots), although some have diverse portfolios (red dots). Other users are located along the diagonal, indicating that they buy and sell in equal quantities. Most of these users transact with a large set of tokens (orange and red dots) and only a few of them are specialized in a given token.

\begin{table}[t]
    \centering
\begin{tabular}{ |c|c|c|c|c|c|c|c| }
 \hline
& Sells & Buys & Distance & Clustering & Eigenvector  & Closeness  \\
 \hline
Diversity & 0.33& 0.35 & -0.16 & -0.08 & -0.07 & -0.01  \\
 \hline
\end{tabular}
 \caption{Regression coefficients of predicting number of transacted tokens by user (diversity) as a function of network properties: number of sells, number of buys, average distance to others on the network, clustering coefficient, eigenvector centrality and closeness centrality. The model $R^2=0.38$.}
\label{tab:regression}
\end{table}

We applied a regression analysis to understand the relationship between portfolio diversity and other network metrics. The regression coefficients are presented in Table \ref{tab:regression}. We observe that diversity of tokens is positively correlated with buys and sells as we showed in Figure \ref{fig:transactions_per_user_scatter} and negatively correlated with distance and clustering. Diverse users are closer than usual to most people, but connect groups that are not connected to each other or to other groups. That topology is observable in the groups of nodes around hubs shown in Figures \ref{fig:visualization} and \ref{fig:visualization_sub}. 


\section{Discussion}
\label{discussion}

The ERC20 transaction network has properties that resemble social networks from other types of human activities and communication media \cite{MORALES20141,doi:10.1063/1.4913758}. For example, Twitter communication networks also present similar hub-spoke structures, where influential users get retweeted by less popular ones. Eventually a set of users retweet more than one source and link these communities, creating the giant component where most of the information flows. The reason behind such similarities is rooted in the complexity of social systems. Cohesive groups are integrated with other groups in the large scale by means of weaker ties \cite{granovetter1977strength}. More generally, the emergence of network giant components can be framed in random network generation models that present phase transitions \cite{erdos1960evolution}. Nodes are initially connected to a few others, generating multiple small communities. These communities are slowly aggregated with each other, until a few edges connect them and a giant component emerges at larger scales.

We have shown that users and tokens have heterogeneous behaviors in terms of the number and diversity of transactions. Diversification and specialization of portfolios have direct consequences on the fragility of strategies to market changes \cite{Taleb2012}. Specialized portfolios are more sensitive to changes in prices and consumer behaviors, as opposed to diversified sets of assets. However, only a few users seem to be performing diversified transactions. These few users connect communities by bridging and exchanging tokens. While their portfolios might be the most robust to market changes, the system could become vulnerable to their failure, miss-behavior or attack. 

In terms of the heterogeneous number of transactions, a large set of users and tokens show very little activity, while a few of them dominate transactions. This seems to be in contrary to the idea of decentralization proposed by crypto-currencies. However, the centralization of resources occurs at the social level and not on the medium to transact (i.e. banks or distributed ledgers). Emergent accumulation of transactions per user is possible in the presence or absence of centralized authorities. It is imperative to bridge social policies and technology in order to create future tokens and means of exchange that controls for the emergence of inequality and create better societies.


\begin{thebibliography}{10}

\bibitem{nakamoto2008bitcoin}
Nakamoto S.
\newblock Bitcoin: A peer-to-peer electronic cash system. 2008;.

\bibitem{Swan:2015:BBN:3006358}
Swan M.
\newblock Blockchain: Blueprint for a New Economy.
\newblock 1st ed. O'Reilly Media, Inc.; 2015.

\bibitem{WEF2018}
Mulligan C, Scott JZ, Warren S, Rangaswami J.
\newblock Blockchain Beyond the Hype A Practical Framework for Business
  Leaders.
\newblock {WEF:} {W}hite {P}aper. 2018;.

\bibitem{Pentland:2008:HST:1450928}
Pentland AS.
\newblock Honest Signals: How They Shape Our World.
\newblock The MIT Press; 2008.

\bibitem{clack2016smart}
Clack CD, Bakshi VA, Braine L.
\newblock Smart contract templates: foundations, design landscape and research
  directions.
\newblock arXiv preprint arXiv:160800771. 2016;.

\bibitem{tokcatalini2018initial}
Catalini C, Gans JS.
\newblock Initial coin offerings and the value of crypto tokens.
\newblock {National Bureau of Economic Research}. 2018;.

\bibitem{tokyoder1983price}
Yoder JD, Adams J, Prince HT.
\newblock The price of a token.
\newblock JPMS: Journal of Political and Military Sociology. 1983;11(2):325.

\bibitem{tokcong2018tokenomics}
Cong LW, Li Y, Wang N.
\newblock Tokenomics: Dynamic adoption and valuation.
\newblock SSRN. 2018;.

\bibitem{DBLP:journals/corr/abs-1807-00955}
Zargham M, Zhang Z, Preciado VM.
\newblock A State-Space Modeling Framework for Engineering Blockchain-Enabled
  Economic Systems.
\newblock CoRR. 2018;abs/1807.00955.

\bibitem{PhysRevE.89.032804}
Battiston F, Nicosia V, Latora V.
\newblock Structural measures for multiplex networks.
\newblock Phys Rev E. 2014;89:032804.
\newblock doi:{10.1103/PhysRevE.89.032804}.

\bibitem{8029379}
{Zheng} Z, {Xie} S, {Dai} H, {Chen} X, {Wang} H.
\newblock An Overview of Blockchain Technology: Architecture, Consensus, and
  Future Trends.
\newblock In: 2017 IEEE International Congress on Big Data (BigData Congress);
  2017. p. 557--564.

\bibitem{Eyal:2018:MEB:3234519.3212998}
Eyal I, Sirer EG.
\newblock Majority is Not Enough: Bitcoin Mining is Vulnerable.
\newblock Commun ACM. 2018;61(7):95--102.
\newblock doi:{10.1145/3212998}.

\bibitem{rahulamathavan2017privacy}
Rahulamathavan Y, Phan RCW, Rajarajan M, Misra S, Kondoz A.
\newblock Privacy-preserving blockchain based IoT ecosystem using
  attribute-based encryption.
\newblock In: 2017 IEEE International Conference on Advanced Networks and
  Telecommunications Systems (ANTS). IEEE; 2017. p. 1--6.

\bibitem{eyal2016bitcoin}
Eyal I, Gencer AE, Sirer EG, Van~Renesse R.
\newblock Bitcoin-ng: A scalable blockchain protocol.
\newblock In: 13th $\{$USENIX$\}$ Symposium on Networked Systems Design and
  Implementation ($\{$NSDI$\}$ 16); 2016. p. 45--59.

\bibitem{yaniv2018}
Somin S, Gordon G, Altshuler Y.
\newblock Network Analysis of {ERC20} Tokens Trading on {E}thereum Blockchain.
\newblock In: Unifying Themes in Complex Systems IX. Cham: Springer; 2018. p.
  439--450.

\bibitem{7796940}
Maesa DDF, Marino A, Ricci L.
\newblock Uncovering the Bitcoin Blockchain: An Analysis of the Full Users
  Graph.
\newblock In: 2016 IEEE International Conference on Data Science and Advanced
  Analytics (DSAA); 2016. p. 537--546.

\bibitem{fi8010007}
Lischke M, Fabian B.
\newblock {A}nalyzing the Bitcoin Network: The First Four Years.
\newblock Future Internet. 2016;8(1).
\newblock doi:{10.3390/fi8010007}.

\bibitem{Krafft2016}
P~M~Krafft WPNDPYAESJBTAP J~Zheng.
\newblock Human collective intelligence as distributed Bayesian inference.
\newblock arXiv. 2016;1608.01987.

\bibitem{Yaniv2014}
Liu Y, Nacher JC, Ochiai T, Martino M, Altshuler Y.
\newblock Prospect Theory for Online Financial Trading.
\newblock PLOS ONE. 2014;9(10):1--7.
\newblock doi:{10.1371/journal.pone.0109458}.

\bibitem{Pentland2012}
Pan W, Altshuler Y, Pentland A.
\newblock Decoding Social Influence and the Wisdom of the Crowd in Financial
  Trading Network.
\newblock In: 2012 International Conference on Privacy, Security, Risk and
  Trust and 2012 International Confernece on Social Computing; 2012. p.
  203--209.

\bibitem{UserBeh2006}
Slob A, Verbeek PP.
\newblock Technology and user behavior.
\newblock Springer Netherlands; 2006.

\bibitem{BarYam2005ComplexityR}
Bar-Yam Y.
\newblock Complexity Rising : from Human Beings to Human Civilization , a
  Complexity Profile; 2005.

\bibitem{Ashby1991}
Ashby WR.
\newblock In: Requisite Variety and Its Implications for the Control of Complex
  Systems. Boston, MA: Springer US; 1991. p. 405--417.

\bibitem{Pompianu2017}
Bartoletti M, Pompianu L.
\newblock An empirical analysis of smart contracts: plat- forms, applications,
  and design patterns.
\newblock International Conference on Financial Cryptography and Data Security.
  2017;.

\bibitem{Anderson2016}
L~Anderson APPR R~Holz, Weber I.
\newblock New kids on the block: an analysis of modern blockchains.
\newblock arXiv. 2016;1606.06530.

\bibitem{Christidis2016}
Christidis K, Devetsikiotis M.
\newblock Blockchains and smart contracts for the internet of things.
\newblock IEEE Access. 2016;4.

\bibitem{CHEN2018567}
Chen Y.
\newblock Blockchain tokens and the potential democratization of
  entrepreneurship and innovation.
\newblock Business Horizons. 2018;61(4):567 -- 575.
\newblock doi:{https://doi.org/10.1016/j.bushor.2018.03.006}.

\bibitem{yaniv20182}
S~Somin APYA G~Gordon.
\newblock Dynamic Equilibration of ERC20 Network.
\newblock CODE Conference. 2018;.

\bibitem{Pentland2014}
Shmueli E, Altshuler Y, Pentland A.
\newblock Temporal Dynamics of Scale-Free Networks.
\newblock In: Social Computing, Behavioral-Cultural Modeling and Prediction.
  Cham: Springer International Publishing; 2014. p. 359--366.

\bibitem{MORALES20141}
Morales AJ, Borondo J, Losada JC, Benito RM.
\newblock Efficiency of human activity on information spreading on Twitter.
\newblock Social Networks. 2014;39:1 -- 11.
\newblock doi:{https://doi.org/10.1016/j.socnet.2014.03.007}.

\bibitem{doi:10.1063/1.4913758}
Morales AJ, Borondo J, Losada JC, Benito RM.
\newblock Measuring political polarization: Twitter shows the two sides of
  Venezuela.
\newblock Chaos: An Interdisciplinary Journal of Nonlinear Science.
  2015;25(3):033114.
\newblock doi:{10.1063/1.4913758}.

\bibitem{granovetter1977strength}
Granovetter M.
\newblock The strength of weak ties.
\newblock In: Social networks. Elsevier; 1977. p. 347--367.

\bibitem{erdos1960evolution}
Erdos P, R{\'e}nyi S.
\newblock On the evolution of random graphs.
\newblock Publ Math Inst Hung Acad Sci. 1960;5(1):17--60.

\bibitem{Taleb2012}
Taleb N.
\newblock Antifragile: {T}hings that gain from disorder.
\newblock Random House; 2012.

\end{thebibliography}
\end{document}